\documentclass[a4paper,10pt,twoside]{cpc-hepnp}

\usepackage{multicol}
\usepackage{graphicx}
\usepackage{booktabs}
\usepackage{amssymb,bm,mathrsfs,bbm,amscd}
\usepackage[tbtags]{amsmath}
\usepackage{lastpage}
\usepackage{CJK}

\begin{document}
\begin{CJK*}{GBK}{song}

\fancyhead[c]{Submitted to `Chinese Physics C'}%\small Chinese Physics C~~~Vol. 37, No. 1 (2013) 010201} 
\fancyfoot[C]{%\small 010201-
\thepage}

\footnotetext[0]{Received 17 July 2013}

\title{A corresponding-state approach to quark-cluster matter\thanks{Supported by the National Basic Research Program of China (2012CB821800, 2009CB824800), the National
Natural Science Foundation of China (11225314, 11203018, 10935001) and
the National Fund for Fostering Talents of Basic Science.}}

\author{%
      GUO Yan-Jun%(¹ùÑåŸý)
      $^{1;1)}$\email{guoyj10@pku.edu.cn}%
\quad LAI Xiao-Yu%(ÀŽÐ¡Óí)
$^{2;2)}$\email{xylai4861@gmail.com}%
\quad XU Ren-Xin%(ÐìÈÊÐÂ)
$^{1;3)}$\email{r.x.xu@pku.edu.cn}%
}

\maketitle

\address{%
$^1$School of Physics and State Key Laboratory of Nuclear Physics and
Technology, Peking University, Beijing 100871, China\\
$^2$ School of Physics, Xinjiang University, Urumqi 830046, China\\
}

\begin{abstract}
The state of super-dense matter is essential for us to understand
the nature of pulsars, but the non-perturbative quantum
chromodynamics (QCD) makes it very difficult for direct calculations
of the state of cold matter at realistic baryon number densities
inside compact stars.
Nevertheless, from an observational point of view, it is conjectured
that pulsars could be made up of quark clusters since the strong
coupling between quarks might render quarks grouped in clusters.
We are trying an effort to find an equation of state of
condensed quark-cluster matter in a phenomenological way.
Supposing that the quark-clusters could be analogized
to inert gases, we apply here the corresponding-state approach to
derive the equation of state of quark-cluster matter, as was
similarly demonstrated for nuclear and neutron-star matter in
1970s.
According to the calculations presented, the quark-cluster stars,
which are composed of quark-cluster matter, could then have high
maximum mass that is consistent with observations and, in turn,
further observations of pulsar mass would also put constraints to
the properties of quark-cluster matter.
Moreover, the melting heat during solid-liquid phase conversion and
the related astrophysical consequences are also briefly discussed.
\end{abstract}



\footnotetext[0]{\hspace*{-3mm}\raisebox{0.3ex}{$\scriptstyle\copyright$}2013
Chinese Physical Society and the Institute of High Energy Physics
of the Chinese Academy of Sciences and the Institute
of Modern Physics of the Chinese Academy of Sciences and IOP Publishing Ltd}%

\begin{multicols}{2}

\section{Introduction}

The state of matter above the nuclear matter density, $\rho_0$, is
still far from certainty, whereas it is essential for us to explore
the nature of compact stars.
At average density higher than $\sim 2\rho_0$, the quark degree of
freedom inside would not be negligible, and historically such
compact stars are called quark
stars~\citep{Itoh:1970uw,Witten:1984rs, Alcock:1986hz, Haensel1986}.
%
%The Bodmer-Witten conjecture~\citep{Bodmer1970, Witten1984}, which states
%that strange quark matter
%(composed of $u$, $d$ and $s$ quarks) could be more stable than nuclear
%matter, makes it possible that pulsars could actually be the so-called quark stars.
%
Although cold quark matter is difficult to be created in laboratories
or studied by direct QCD calculations, some efforts have been made
to model quark matter and quark stars, from MIT bag model to color
super-conductivity model~\citep{Alford2008}.
In most of these models, quark matter could usually be characterized
by soft equation of state, because the asymptotic freedom of QCD
tells us that as energy scale goes higher, the interaction between
quarks becomes weaker.
Nonetheless, astrophysical phenomenology of compact stars can still
not rule out that striking physical possibility, making pulsars as
superb astrophysical laboratories~\citep[see e.g.][and references
therein]{Weber1999, Weber2005}.

However, at realistic baryon densities of compact stars, $\rho \sim
(2-10)\rho_0$, the energy scale is usually below 0.8 GeV, which is
much lower than the scale where the asymptotic freedom could apply.
In contrast, the non-perturbative effect should be significant, making
quarks to couple strongly with each other.
Quark-clustering is then conjectured to occur in cold dense matter
inside compact stars, by condensation of quarks in position space
due to the strong coupling between quarks~\citep{Xu03}.
A realistic quark star could then be actually a ``quark-cluster
star'', and solidification could be a natural result if the kinetic
energy of quark-clusters is much lower than the residual interaction
energy between the clusters.
The idea of clustering quark matter could provide us a way to
understand different manifestations of pulsar-like compact
stars~\citep{Xu:2010, LX10C}.

How can then one model the equation of sate of quark-cluster matter?
Due to the lack of both theoretical and experimental evidence, the
hypothetical quark-clusters in cold dense matter have not been
confirmed, and it is also difficult for us to derive the properties
of quark-cluster matter from the first principles of QCD
calculations.
Nevertheless, an empirical way was employed and discussed seriously
in a {\it corresponding-state approach} to deduce the properties of
nuclear matter in 1970s~\citep[e.g.][]{PA1974,Canuto1975}.
To establish a model which could be tested by observations, we adopt
an empirical method here, by analogizing quark-clusters to inert
gases and applying the corresponding-state approach too.
A quark-cluster is usually assumed to be colorless, just like an
inert atom being electric neutral.
The interaction between inert gas atoms is the result of residual
electromagnetic force, and similarly the interaction of
quark-clusters could be regarded as the result of residual strong
force, both of which should be characterized by the short-distance
repulsion and long-distance attraction.
In this paper, we assume that the interaction between quark-clusters
could be described approximately by the same form as that between
inert gas atoms, i.e. the Lennard-Jones potential, only with
different parameters indicating stronger interaction and larger
densities.

In fact, quark matter in Lennard-Jones model has been studied,
where the equation of state is derived via summing the interaction
energy of all quark-clusters~\citep{LX09b}.
Previously, a polytropic model~\citep{LX09} and a two-Gaussian
component soft-core model~\citep{NX11} for quark-cluster stars
has also been applied.
The so-called corresponding-state approach we demonstrated in this
paper, however, is an empirical one, to derive properties of
quark-cluster matter by just a comparison to the experimental data
of inert gases, based on the law of corresponding states.
The law of corresponding states was first proposed by de
Boer~\citep{deBoer1948a}, who found that the properties of inert
gases, such as pressure and density, could be written in a reduced
form.
After reducing to dimensionless terms, the experimental data of
various inter gases can be fitted in smooth curves with a single
quantum parameter.
If the quark-cluster matter is assumed to be similar to inert gases,
the corresponding-state approach can also be applicable to study the
state of quark-cluster matter, without knowing its exact structure.

With a similar form of interaction, we may derive the equation of
state of quark-cluster matter from empirical data of inert gases,
through the corresponding-state approach.
The masses and radii of quark-cluster stars can then be derived and
compared with observations.
We find that the maximum mass of quark-cluster stars can be well
above $2M_\odot$.
Although in principle one may obtain a maximum mass high enough to
explain observations in any kinds of unphysical models, we call
attention that the quark-cluster star model has meaningful
implications for one to understand different manifestations, e.g.,
of the surface \citep{DX2012}.
Additionally, the melting heat is also discussed, and it is shown
that the solidification of newly born quark-cluster stars might
explain the plateau of $\gamma$-ray bursts.

This paper is organized as following.
We summarize the properties and observational implications of
quark-cluster stars in \S2, and a brief introduction of the law of
corresponding states is given in \S3.
The equation of state of quark-cluster matter and the mass-radius
curve of quark-cluster stars are derived in \S4 using the
corresponding-state approach.
The melting heat of solid quark-cluster stars and the related
astrophysical consequences are discussed in \S5.
We make conclusions and discussions in \S6.

\section{Quark-cluster matter}

A lot of basic intuition questions are frequently asked about
quark-cluster matter though such a state was proposed ten years
ago~\citep{Xu03}.
Would quark-cluster matter be more energetically favored than nuclear
matter or strange quark matter?
Can quark-clusters be analogous to inert gases?
Could quark-cluster star model be really necessary in astrophysics
of compact stars?
Certainly we cannot present clear and final answers to these because
of both micro- and astro-problems as well as their entanglement.
In order for readers to have a thorough and comprehensive view of
the quark-cluster star idea, we here make some rough estimation
about quark-cluster existence, and demonstrate that the answers
could be positive in some region of the parameter space in QCD phase
diagram.
Observational hints are also summarized.

\subsection{The stability of quark-cluster matter}

It's an interesting but difficult problem to know the stability from
first principle.
A special kind of quark-cluster, so-called H-cluster, was studied extensively,
and by comparing energy per baryon at fixed density, it's found that H-cluster
matter might be more stable than both neutron matter and nuclear matter
when the density is larger than 2$\rho_0$, where the in-medium effect plays
the crucial role in stabilizing H-cluster matter \citep[see][Sect. 2]{LGX12}.
Besides, an order of magnitude estimation could also help compare the
stability of these three states.

{\it Nuclear matter vs. Quark-cluster matter.}
In the low energy region of QCD phase diagram, quarks are confined
in nucleons.
However, at the density of realistic compact stars, the confined state
may not be simply that of hadrons, because a light-flavor symmetry
is likely to be restored.
In ordinary case, electrons are outside the nucleus and their energy
$E_e$ is far less than 1 MeV, so atoms could be stable with 2-flavor
symmetry.
Nevertheless, things are different in case of pulsar, for that
electrons are inside the gigantic nucleus and the Fermi energy of
electrons would be $E_e \sim 10^2$ MeV, even larger than the mass
difference between $s$ quark and $u/d$ quark.
Such a high energy might intensify the interaction $e+p \to
n+\nu_e$, thus $E_e$ decreases but the nuclear symmetry energy
increases.
Therefore $s$ quark is likely to be excited in gigantic nucleus, the
number of which may be slightly ($\sim 10^{-5}$) less than $u/d$
quark as $s$ quark is heavier.
If 3-flavor symmetry is restored, the number of electrons in pulsar would
be much less, which makes $E_e \sim 10$ MeV, and the gigantic nucleus
would be stable.
Furthermore, three flavors of quarks could be grouped together to
form a new hadron-like confined (quark-cluster) state in gigantic
nucleus if the coupling between quarks is still strong.

{\it Strange quark matter vs. Quark-cluster matter.}
At the high density, low temperature regime, cold dense quark matter could
be of Fermi gas or liquid if the interaction between quarks is negligible.
However, the questions is: can the density in realistic compact stars be
so high that we can neglect the interaction?
The average density of a pulsar-like star with typical mass of 1.4
$M_\odot$ and radius of 10 km is only $\sim 2.4 \rho_0$.
For 3-flavor quark matter with density of 3$\rho_0$, we have number
densities for each flavor of quark, $u$, $d$, and $s$, of $n_u
\simeq n_d \simeq n_s \sim(3 \times 0.16 = 0.48) \textrm{ fm}^{-3}
$.
A further calculation of Fermi energy gives,
$E_F^{NR} \approx \frac{\hbar^2}{2m_q} (3\pi^2)^{2/3} \cdot n^{2/3} =380 \textrm{ MeV}$
if quarks are considered moving non-relativistically, or
$E_F^{ER} \approx \hbar c (3\pi^2)^{1/3} \cdot n^{1/3} = 480 \textrm{ MeV}$
if quarks are considered moving extremely relativistically.

However, the interaction between quarks may play an important role
in determining the real state.
For a quark with length scale $l$, from Heisenberg's uncertainty relation,
the kinetic energy would be of $\sim p^2/m_q \sim \hbar^2/(m_q l^2)$,
which has to be comparable to the color interaction energy of $E\sim \alpha_s \hbar c/l$
in order to have a bound state, where $\alpha_s$ is the coupling constant
of strong interaction.
One then finds if quarks are dressed, with a mass of $m_q=300$ MeV,
\begin{equation}\label{}
 l \sim \frac{1}{\alpha_s} \frac{\hbar c}{m_q c^2} \simeq \frac{1}{\alpha_s} \textrm{ fm, }
 E \sim \alpha_s^2 m_q c^2 \simeq 300 \alpha_s^2 \textrm{ MeV.}
\end{equation}
This is dangerous for the Fermi state of matter since $E$ is
approaching and even greater than the Fermi energy of $\sim 0.4$ GeV
if the running coupling constant $\alpha_s >$  1, and a Dyson-Schwinger
equation approach to non-perturbative QCD shows that the color
coupling should be very strong rather weak, with $\alpha_s \gtrsim 2$
at a few nuclear densities in compact stars \citep{Xu:2010}.
Such strong interaction could render the quark grouped into
clusters, rather than condensation in momentum space to form a color
super-conductivity state.

%The estimate above is made in case of 3$\rho_0$, about the average
%density of a typical pulsar, and it may be doubted would things be different
%in the central region of star.
%
%However, such a density could be representative since the central density of
%quark-cluster star is just a few times the surface density, usually less than
%10$\rho_0$, due to strong self-bounding.

\subsection{Properties of quark-cluster matter}

From arguments above, we can see that a symmetry of light flavor quarks
is restored in the quark-clustering phase, which
is different from the usual hadron phase. The color interaction is still
strong there.
On the other hand, the quark-clustering phase is also different from the
conventional quark matter phase which is composed of relativistic and
weakly interacting quarks.
The quark-clustering phase could thus be considered as an intermediate state
between hadron phase and free-quark phase.

Compact stars composed of pure quark-clusters are electric neutral, but
in reality there could be some flavor symmetry breaking that leads to the
non-equality among $u$, $d$ and $s$, usually with less $s$ than $u$ and $d$.
The positively charged quark matter is necessary because it allows the
existence of electrons that might be crucial for us to understand the radiation
behaviors of compact stars.

What could be a realistic quark-cluster?
We know that $\Lambda$ particles (with structure $uds$) possess light-flavor
symmetry, and one may think that a kind of quark clusters would be
$\Lambda$-like.
However, the interaction between $\Lambda$ is attractive, so H-cluster
with structure $uuddss$ could emerge, which was previously predicted to be a stale
state or resonance~\citep{Jaffe1977}, and recently Lattice QCD simulations
have shown possible evidence for its existence~\citep{Beane:2010hg, Inoue:2010es}.
Besides H-cluster, an 18-quark cluster, i.e. quark-$\alpha$, being completely symmetric
in spin, color and flavor space, was also speculated to exist~\citep{Michel1988}.
The number of quarks in one cluster is left as a free parameter in this paper,
and we set $N_q=6$ and $N_q=18$ for the sake of simplicity in the calculations
as following, corresponding to $H$-cluster and quark-$\alpha$, respectively.

An estimate for the length scale $l_{qc}$ of quark-cluster gives
$ l_{qc} \sim  1/\alpha_s \textrm{ fm} \lesssim 1$ fm, which would
be less than the average distance between quark-clusters
$d\simeq(3\times 0.16/N_q)^{-1/3} \gtrsim 1$ fm.
Although quark-clusters consist of more quarks, they might not be larger than
nucleons, and the distance between quark-clusters would be larger, so it is not
likely that quark-clusters would be in closer proximity than nucleons.
The quantum effect would not be significant if the residual short-distant
repulsing interaction works, and quark-cluster can be considered as classical
particles rather than that of quantum gas.
What's more, quark-cluster may move non-relativistically due to large mass
and could be localized in lattice at low temperature.

When justifying the corresponding-state approach, one may doubt why
should the law of corresponding states apply to quarks.
Theoretically, the corresponding-state law reflects the statistical
behavior of system. What's required is just the same form of interaction
potential, while certain values would not influence the conclusion if
the parameters are re-scaled.
Then another question may be: could the interaction between quark-clusters
be described by Lennard-Jones potential?
It's hard to know the accurate form of the interaction between quark-clusters,
but it could have a similar shape as Lennard-Jones potential, considering the
property of short-distance repulsion and long-distance attraction.
Quark-cluster could be analogized to nucleons, except for light-flavor
symmetry.
Since strong interaction is not sensitive to flavor, the interaction
between quark-clusters should be similar to that of nucleons, which is found
to be Lennard-Jones-like by both experiment and modeling~\citep{Wilczeck2007}.

The interaction between quark-clusters may not be perfectly described by
Lennard-Jones potential, the long range part of which may be proportional
to $r^{-7}$ instead of $r^{-6}$ ~\citep{Fujii1999}.
As a zeroth approximation, the Lennard-Jones potential assumption may
lead to violation of the law of corresponding states, but the reduced properties
of quark-cluster matter should at least be in the same range with, even not
exactly fall in the experimental lines of inert gases.
So our approach is to some degrees reasonable, when the exact approach
under QCD calculations seems to be impossible due to the significant
non-perturbative effect.

\subsection{Observational hints for the nature of pulsar}

In the absence of QCD calculations due to non-perturbative effect,
estimations in \S2.1 could only demonstrate the possibility of
stable quark-cluster matter, while no further conclusion could be
made and the stability of quark-cluster remains to be justified.
Nevertheless, quark-cluster was speculated to exist primarily for understanding
astrophysical observations of pulsar-like compact stars.
In addition to first principles, pulsar-like compact stars could also provide
valuable information for properties of super-dense matter, different manifestations
of which provide hints of the state of matter at supra-nuclear density.
Various observational phenomena could be understood in terms of quark-cluster
star model, including those that are challenging in conventional neutron star models
~\citep{SQM2008}.

What if pulsar is made of quark-cluster matter?
There are at least three consequences relevant to observational phenomena.

{\it A stiff equation of state.}
It is conventionally thought that the state of dense matter softens and thus
cannot result in high maximum mass if pulsars are quark stars, and that the
discovery of 2$M_\odot$ pulsar PSR J1614-2230 \citep{Demorest:2010bx}
may make pulsars unlikely to be quark stars.
However, quark-cluster star would have a stiff equation of state,
because quark-cluster should be non-relativistic particle for its
large mass, and because there could be strong short-distance
repulsion between quark-clusters.
It may well be possible to obtain a maximum mass of  $\geqslant 2M_\odot$.
Certainly, finding a stiffer equation of state is not enough to
claim quark-cluster matter exist, but other observations may hints
for a self-bound surface and global solid structure, which could
also favor the existence of quark-cluster.

{\it A self-bound surface.}
Different from traditional neutron stars, quark-cluster star would
be self-bound by residual color-interaction between clusters, which
could be a crucial difference providing observational manifestations
to distinguish the two models.

Drifting subpulses phenomena in radio pulsars suggest the existence
of Ruderman-Sutherland-like gap-sparking and thus strong binding of
particles on pulsar polar caps to form vacuum gaps, but the
calculated binding energy in normal neutron star models could not be
so high unless the magnetic field is extremely strong.
This problem could be naturally solved in quark-cluster star scenario due to
the strong self bound nature on surface \citep{Xu1999, Qiao04}.

In addition, many theoretical calculations predict the existence of
atomic features in the thermal X-ray emission of neutron star
atmospheres, while none of the expected spectral features has been
detected with certainty up to now, which hints that there might not
exist the atmospheres speculated in conventional models.
Though modified neutron star atmospheres with very strong surface
magnetic fields \citep{HL03, Turolla04} might reproduce a
featureless spectrum too, a natural suggestion to understand the
general observation could be that pulsars are actually quark-cluster
star without atoms on the surface \citep{Xu02}.

In addition, the bare and chromatically confined surface of
quark-cluster star could overcome the baryon contamination problem
and create a clean fireball for $\gamma$-ray burst and supernova.
The strong surface binding would result in extremely energetic exploding
because the photon/lepton luminosity of a quark-cluster surface is not limited
by the Eddington limit, and supernova and $\gamma$-ray bursts could then
be photon/lepton-driven \citep{Ouyed05, Paczynski05, Chen07}.
Recently, it was shown that the magnetic field observed for some
compact stars could be generated by small amounts of differential
rotation between the quark matter core and the electron sea~\citep{Negreiros12}.

{\it A global solid structure.}
Quark-cluster star could be in a global solid state, like ``cooked
eggs'', if the kinetic energy is less than interaction energy
between quark-clusters, while for normal neutron stars, only crust
is solid, like ``raw'' eggs.
Rigid body would precess naturally when spinning, either freely or
by torque, and the observation of possible precession or even free
precession of B1821-11 \citep{Stairs2000} and others could suggest a
global solid structure of pulsars.

Star-quake is a peculiar action of solid compact stars, during which huge
free energy, such as gravitational and elastic energy, would be released.
For a pulsar with
mass $M\sim M_\odot$ and radius $R\sim 10$ km, the stored gravitational
energy is $\simeq GM^2/R \sim 10^{53}$ erg, so  energy released would be
$\sim 10^{53} \Delta R/R$ when the radius changes from $R$ to $R-\Delta R$.
Compared with magnetars powered by magnetic energy, quake-induced
energy in solid quark-cluster stars may also be enough to power the
bursts, flares and even superflares of soft $\gamma$-ray repeaters
and anomalous X-ray pulsars \citep{Xu06}.

Combining surface and global properties, we think that the
quark-cluster star model would be reasonable to describe pulsar-like
stars, while this paper is mainly focused on the equation of state
for quark-cluster stars, via a phenomenological way.

\section{The law of corresponding states}

The law of corresponding states, advocated by de Boer~\citep{deBoer1948a},
shows that the equation of state of substances with same form of interaction
can be written in a reduced and universal form.
Consider a group of substances with the following properties:
(1) the total potential energy due to interaction can be written as a sum of
identical expressions $\varphi(r_{ik})$, each of which depends only on the
distance $r_{ik}$ between two particles $i$ and $k$;
(2) $\varphi(r)=\varepsilon f(r/\sigma)$, where $f$ is a function same for
all substances, and $\varepsilon$, $\sigma$ are characteristic energy and
length for different species.
The macroscopic quantities, such as pressure $P$, volume $V$ and
temperature $T$, can be expressed in dimensionless terms:
\begin{eqnarray}
  P^* &=& P\sigma^3/\varepsilon \label{eq-p} \\
  V^* &=& V/(N\sigma^3) \label{eq-v} \\
  T^* &=& kT/\varepsilon
\end{eqnarray}
Another dimensionless parameter is
\begin{equation}\label{eq-l}
  \Lambda^*=h/(\sigma\sqrt{m\varepsilon}),
\end{equation}
corresponding to the de Broglie wavelength, which is constructed to measure
the importance of quantum effects.
It can be proved that the reduced equation of states expressed in dimensionless
quantities is a universal relation
\begin{equation} \label{eq-pf}
P^*=f(V^*,T^*,\Lambda^*),
\end{equation}
which is the formulation of the law of corresponding states~\citep{deBoer1948a}.

Despite the so-called universal equation of states is just formally written
as Eq.~(\ref{eq-pf}), a formula that is difficult to be derived theoretically for most
cases, it could be used to obtain information on the equation of state of a
substance which we are unfamiliar with.
For determined $V^*$ and $T^*$, $P^*$ depends on the value of $\Lambda^*$,
and the $P^*-\Lambda^*$ curve can be drawn using experimental data of
laboratory substances.
If the curve is smooth enough, the value of $P^*$ for unfamiliar matter at
such a state can be predicted provided its $\Lambda^*$ is known.

For some substances described by Lennard-Jones 6-12 potential
\begin{equation}
  \varphi(r)=\varepsilon\{\frac{4}{(r/\sigma)^{12}}-\frac{4}{(r/\sigma)^6}\},
\end{equation}
deBoer had determined $\varepsilon$ and $\sigma$ of noble gases
and some permanent gases ($r=\sigma$ is the distance where $\varphi(r)=0$,
and $\varepsilon$ is the depth of potential well)~\citep{deBoer1948a}.
Then the experimental data of $P^*,T^*$ or $V^*$ for different substances
turn out to be smooth functions of $\Lambda^*$ as corresponding states.
In Fig.~\ref{f-v}, experimental data of the volume $V_0$ at zero temperature
and zero pressure, reduced to $V^*_0=V_0/(N\sigma^3)$, are plotted with
$\Lambda^*$ for some substances.
\begin{center}
  %\centering
   %Requires \usepackage{graphicx}
  \includegraphics[width=2.5 in]{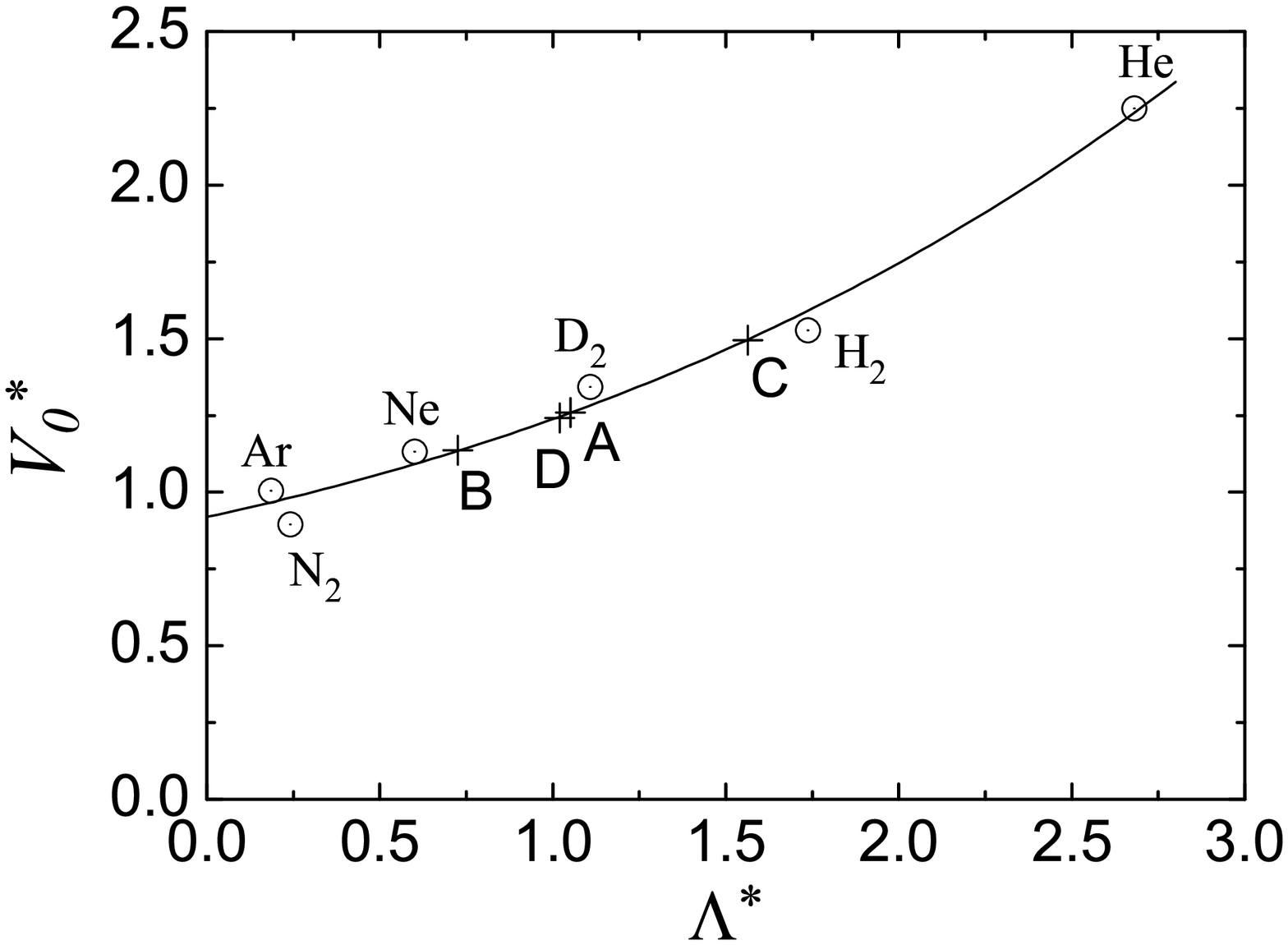}\\
  \figcaption{Experimental data of reduced volume $V_0^*$ and $\Lambda^*$
  at zero temperature and pressure for different inert gases are
  shown by dots~\citep{deBoer1948a}, and the fitted curve of Eq.~(\ref{fit})
  is shown by solid line. Four cases A, B, C, and D are denoted by crosses,
  which will be explained in Sect. 4.1. }\label{f-v}
\end{center}
A smooth curve can be drawn by fitting all the points,
which forms the bases of our prediction via corresponding states law,
and the formula for the fitted curve is
\begin{equation}\label{fit}
  V^*_0=0.57+9.45\times10^{-5}(\Lambda^*+6.35)^{4.44}.
\end{equation}

Considering the property of short-distance repulsion and long-distance
attraction shown by Lennard-Jones potential, we assume that the interaction
between quark-clusters can also be described by this form.
The distinctions between quark-cluster matter and ordinary substances
should be a much deeper
potential well (larger $\varepsilon$) and higher density (smaller $\sigma$).
With the same form of interaction as that of inert gas, we could apply the
law of corresponding states to derive the properties of quark-cluster matter.
If we find the quantum parameter $\Lambda^*$ corresponding to quark-cluster
matter, then $V^*_0$ and other properties that vary smoothly with $\Lambda^*$
can be determined by simply looking at the experimental curves of that property
vs $\Lambda^*$ for inert gases.
%%%%%%%%%%%%%%%%%%

%
\section{The state of quark cluster matter}
\subsection{Parameters}
To apply the law of corresponding states to quark-cluster matter,
we must determine $\varepsilon,\sigma$ and the mass $m$ of each
quark-cluster first.
$m$ depends on the number of quarks $N_q$ and the mass of each quark
$m_0$ in one cluster.
We give each quark a constituent mass and assume $m_0$ is one-third of the
nuclear mass.
$N_q$ is left as a free parameter in this paper, and we set $N_q=6$ and $N_q=18$
for our calculations, corresponding to $H$-cluster and quark-$\alpha$ respectively.

As no experimental attempt has been made to get the values of
$\varepsilon$ and $\sigma$, we try to constrain their values by the surface
density $\rho_s$ of quark-cluster stars.
The temperature of quark stars can be approximated to be zero,
and the pressure also reaches zero at the surface of stars.
Given the value of $V^*_0$, we can calculate the surface density $\rho_s$
(rest-mass density).
It is obvious that $\rho_s$ can be written as
\begin{equation} \label{eq-r1}
\rho_s=N\cdot N_q m_0/V_0,
\end{equation}
and comparing Eq.~(\ref{eq-v}) with Eq.~(\ref{eq-r1}) we can get
\begin{equation} \label{eq-r2}
\rho_s=N_q m_0/(V^*_0\sigma^3).
\end{equation}
For certain values of $N_q$, $\varepsilon$ and $\sigma$,
we can calculate $\Lambda^*$ of quark cluster matter by Eq.~(\ref{eq-l}),
and $V^*_0$ can be found according to the fitted relation Eq.~(\ref{fit}) of
$V^*_0$-$\Lambda^*$ curve,
then we may determine $\rho_s$ using Eq.~(\ref{eq-r2}).
In Fig.~\ref{f-rho}, pairs of $\varepsilon$ and $\sigma$ that correspond to the
same surface density $\rho_s$ are plotted respectively for $N_q=6$ and
$N_q=18$, where values of $\rho_s$ are chosen to be once, twice and three
times of nuclear matter density $\rho_0$.
The lines of $\varepsilon$ and $\sigma$ giving the same
$\Lambda^*$ with values 1, 2 and 3 are also drawn here for a further limit.

The surface density of quark stars is assumed to be in the range
$1<\rho_s/\rho_0<3$.
Quark-clusters could condensate to form solid state like classical particles,
so the quantum effects may not be large for quark-cluster matter, then
$\Lambda^*$ should satisfy $\Lambda^*<2$.
We select four points numbered A, B, C and D representatively to
deduce the equation of state for quark-cluster matter and then
the mass-radius relation of quark-cluster star.
The values of $\varepsilon$, $\sigma$ and the resulting $\rho_s$,
$\Lambda^*$ at points A to D are given in
Table ~\ref{tab1}.
They are also plotted in Fig.~\ref{f-v}, corresponding to four
different cases for quark-cluster matter, in which
the equation of state will be calculated respectively.

\begin{center}
%Table~\ref{table1}
\tabcaption{ \label{tab1} }
\footnotesize

%\vspace{0.5 cm}

%\begin{tabular}{80mm}{c|c|c|c|c|c}
\begin{tabular*}{80mm}{c@{\extracolsep{\fill}}ccccc}

%\tabcolsep 0pt
%\caption{Table}
\toprule
 \ \ \ \ \ & \ $N_q$ & $\varepsilon$ (MeV) & $\sigma$ (fm) & \ \ $\Lambda^*$ \ \  & \ $\rho_s/\rho_0$ \ \\\hline
 A & 18 & 40 & 2.5 & 1.05 & 1.87  \\%\hline
 B & 18 & 100 & 2.3 & 0.72 & 2.72  \\%\hline
 C & 6 & 150 & 1.5 & 1.56 & 2.47 \\%\hline
 D & 6 & 200 & 2.0 & 1.02 & 1.23 \\
 \bottomrule
 \label{table1}
 \end{tabular*}
 \end{center}

% \vspace{0.5 cm}

%A, $N_q=18, \varepsilon=40\textrm{MeV}, \sigma=2.5$ fm;
%B, $N_q=18, \varepsilon=100$ MeV, $\sigma=2.3$ fm;
%C, $N_q=6, \varepsilon=150$ MeV, $\sigma=1.5$ fm;
%D, $N_q=6, \varepsilon=200$ MeV, $\sigma=2.0$ fm.
%

Our choice of parameters $\varepsilon$ and $\sigma$ comes
from the following consideration.
The depth of potential well for nuclear matter is about 100 MeV,
so it may be reasonable that $\varepsilon= \mathcal{O}$(100 MeV).
The average inter-cluster distance $d$ at the surface of quark-cluster stars
is given by
\begin{equation}
  d=[\frac{3\times0.16(\rho_s/\rho_0)}{N_q}]^{-\frac{1}{3}}.
\end{equation}
With $\rho_s/\rho_0=2$, we get $d=1.84$ fm for $N_q=6$ and $d=2.66$ fm
for $N_q=18$.
Then $\sigma =\mathcal{O}$(1 fm) as it should have the same order of
magnitude as $d$.
It can be seen that the selected parameters are consistent with the
above estimation.
%\begin{tabular}{|c|p{1.5cm}|r|r|r|r|}
%  \hline
%    & $N_q$ & $\varepsilon$/MeV & $\sigma$/fm & $\Lambda^*$ & $\rho_s/\rho_n$ \\ \hline
%  A & 18 & 40 & 2.5 & 1.05 & 1.87 \\ \hline
%  B & 18 & 70 & 3.0 & 0.66 & 1.23 \\ \hline
%  C & 6 & 70 & 1.8 & 1.91 & 1.26 \\ \hline
%  D & 6 & 200 & 2.0 & 1.02 & 1.23 \\
%  \hline
%\end{tabular}
%
\begin{center}
  %\centering
  \includegraphics[width=2.5 in]{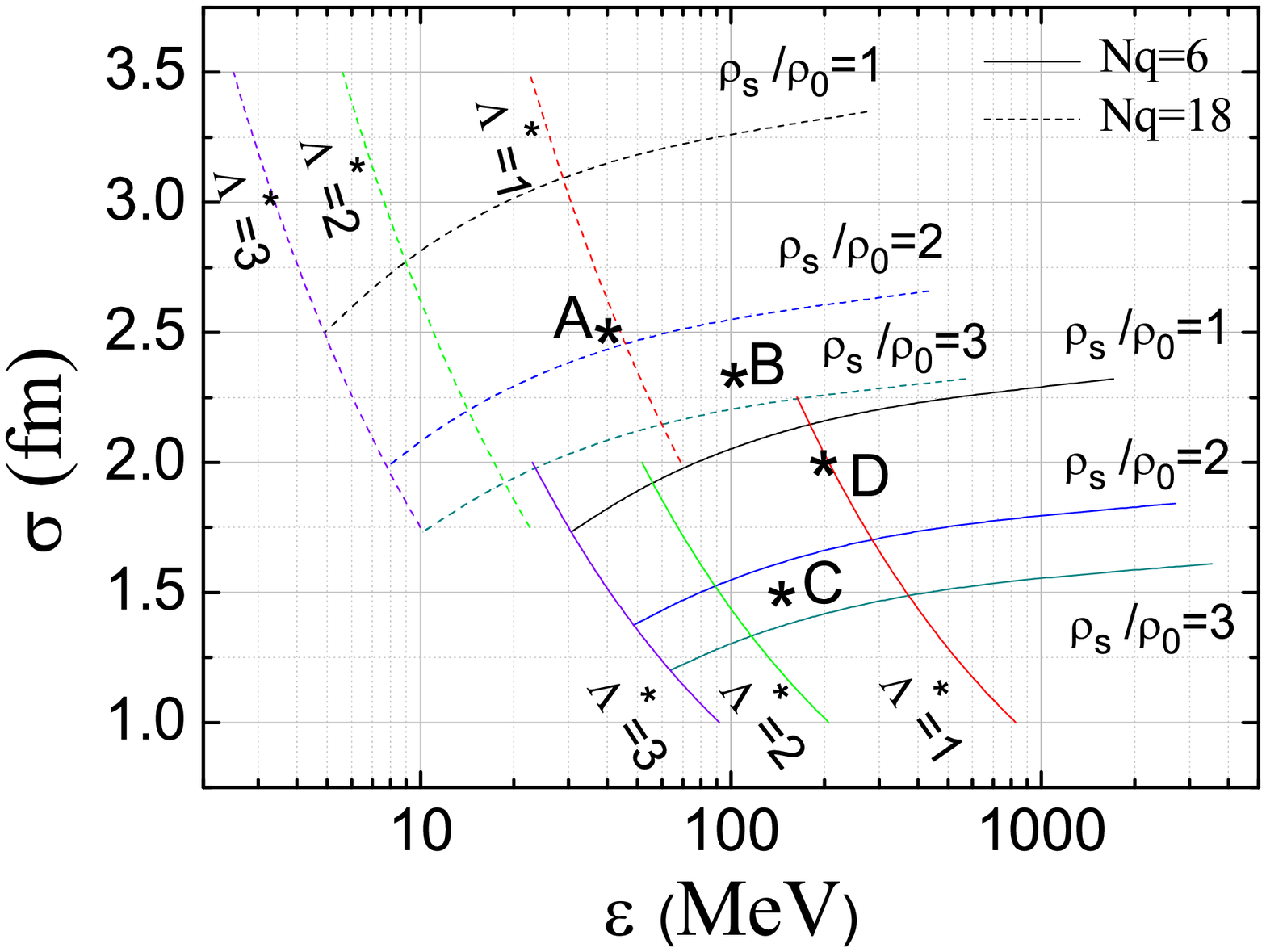}\\
  \figcaption{
  Contour lines of surface density $\rho_s$ and $\Lambda^*$, with solid lines
  representing $N_q=6$ and the dashed lines representing $N_q=18$,
  including $\rho_s/\rho_0=1, 2, 3$ and $\Lambda^*=1, 2, 3$. Four cases
  A, B, C and D are denoted by stars.}\label{f-rho}
\end{center}

\subsection{The equation of state}
Given $\varepsilon$ and $\sigma$, we can deduce the state of quark-cluster
matter by a corresponding-state approach, in the zero temperature case.
If we know the experimental $P^*-\Lambda^*$ curve at a certain $V^*$
and zero temperature, we can find the value of $P^*$ corresponding to
$\Lambda^*$ of quark cluster.
According to Eq.~(\ref{eq-p}) and the number density of quark-clusters
$n=1/(V^*\sigma^3)$, the reduced quantities
$P^*$ and $V^*$ can be converted to $P$ and $n$, then we can get the pressure
at a certain number density of quark-clusters.
Combining this with the relation between mass density $\rho$ (rest-mass density
plus interaction energy density) and number density $n$ of quark-cluster matter,
the equation of state can be derived.

To draw the $P^*-\Lambda^*$ curve at different $V^*$, we need to
know the relationship between $P^*$ and $V^*$ of some substances
at zero temperature.
For the lack of new data, we just use the data provided by de Boer
in his subsequent article~\citep{deBoer1948b}, where values of $P^*$ and
$\Lambda^*$ were given corresponding to different values of $V^*$
for various inert gases.
Taking $V^*=0.88$ for instance, the $P^*-\Lambda^*$ curve are shown in
Fig.~\ref{f-p1}.
The data points are almost in linear relation, which makes our interpolation
reliable.
The value of $\Lambda^*$ at point A is about 1.05, then we find
$P^*\approx25$ from the $P^*-\Lambda^*$ curve.
The corresponding $P$ and $n$ to the reduced quantities $P^*$
and $V^*$ are $P=1.0 \times 10^{35}\textrm{ dyn/cm}^2$, $n/n_0=2.7$
($n_0$ is the number density of nucleons in nuclear matter).
Thus we get $P(n=2.7n_0)=1.0\times10^{35}\textrm{ dyn/cm}^2$ for
quark-cluster matter in case A.
By taking different values of $V^*$, pressure
$P$ at different densities can be determined in case A.
The same procedure is also applicable to the other three cases.

\begin{center}
  %\centering
  \includegraphics[width=2.5 in]{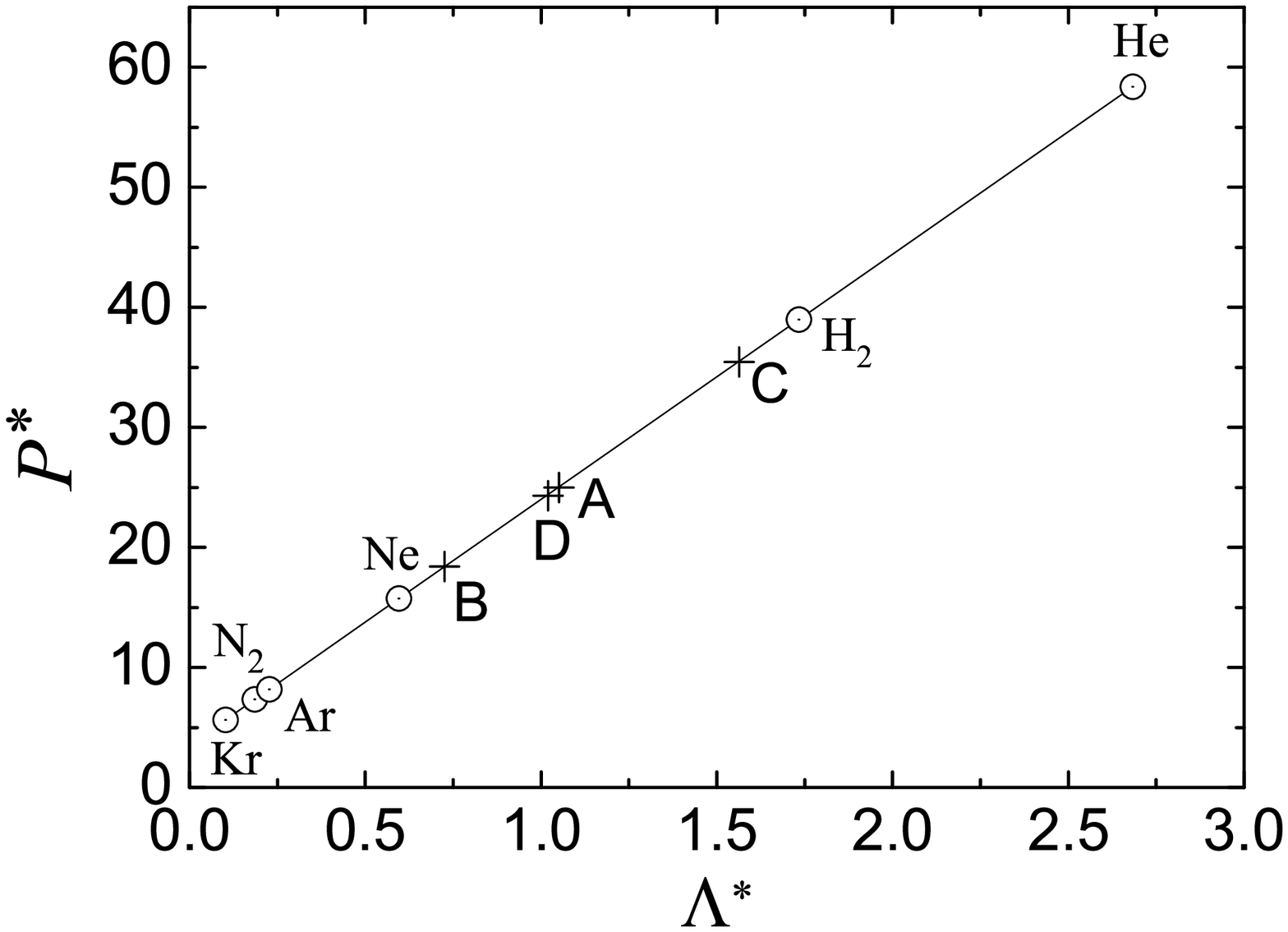}\\
  \figcaption{Experimental data of $P^*$ and $\Lambda^*$
  at zero temperature when $V^*=0.88$ are shown by
  dots~\citep{deBoer1948b}, and the fitted curve is almost a straight
  line (solid line). The cases A, B, C, and D are denoted by crosses.}
  \label{f-p1}
\end{center}

For each set of parameters, what we get is just a set of points in $P-n$ diagram
and not an analytic equation, and then we perform the curve fitting to get an
approximate formula.
The $P-n$ relations derived from curve fitting are
\begin{eqnarray}
  P &=& (2.99\times10^{41}n^{5.63}-1.60\times10^{34}) \textrm{ dyn/cm}^2 \\
  P &=& (1.99\times10^{41}n^{5.64}-7.63\times10^{34}) \textrm{ dyn/cm}^2 \\
  P &=& (8.10\times10^{38}n^{5.24}-1.69\times10^{35}) \textrm{ dyn/cm}^2 \\
  P &=& (6.69\times10^{40}n^{5.63}-1.63\times10^{35}) \textrm{ dyn/cm}^2
\end{eqnarray}
for A, B, C and D respectively, where $n$ is in units of clusters/fm$^3$.
Certainly it is better to deduce the equation of state from a
border range of densities, making the extrapolation to be more accurate.
Nevertheless, lacking in experimental data of laboratory substances,
we can only make such an approximation at this stage.
It is worth mentioning that the approximation will not have much
influence on the following calculations of the mass-radius curves.

According to $P=n^2\frac{dE}{dn}$, where $E$ is the internal energy
per cluster, we can get
\begin{equation}\label{eq-E1}
  E(n)-E(n_s)=\int_{n_s}^n \frac{P(n)}{n^2} dn,
\end{equation}
where $n_s$ is the number density of quark-clusters on the surface of stars.
We may determine the value of $E(n_s)$ from a corresponding-state point of view,
and then the relation between $E$ and $n$ can be derived by the above integral.
Similar to $V_0^*$, $U_0^*=U_0/(N\varepsilon)$ can be approximated as a smooth
function of $\Lambda^*$, where $U_0$ is the internal energy at zero temperature
and zero pressure.
From the data of laboratory substances~\citep{deBoer1948a}, we derive a fitted
formula for $U_0^*$,
\begin{equation}\label{eq-U}
  U_0^*=-8.72+4.91\Lambda^*-0.71\Lambda^{*2},
\end{equation}
and $E(n_s)$ is thus
\begin{equation}\label{eq-E2}
  E(n_s)=U_0/N=U_0^*\varepsilon.
\end{equation}
As both $P(n)$ and $E(n_s)$ are known, it is able to calculate $E(n)$ from
Eq.~(\ref{eq-E1}).
The results are plotted in Fig.~\ref{f-E}, for four groups of parameters A to D,
and we can see that the internal energy
can be comparable to rest-mass energy at some densities.
\begin{center}
  %\centering
  \includegraphics[width=2.2in]{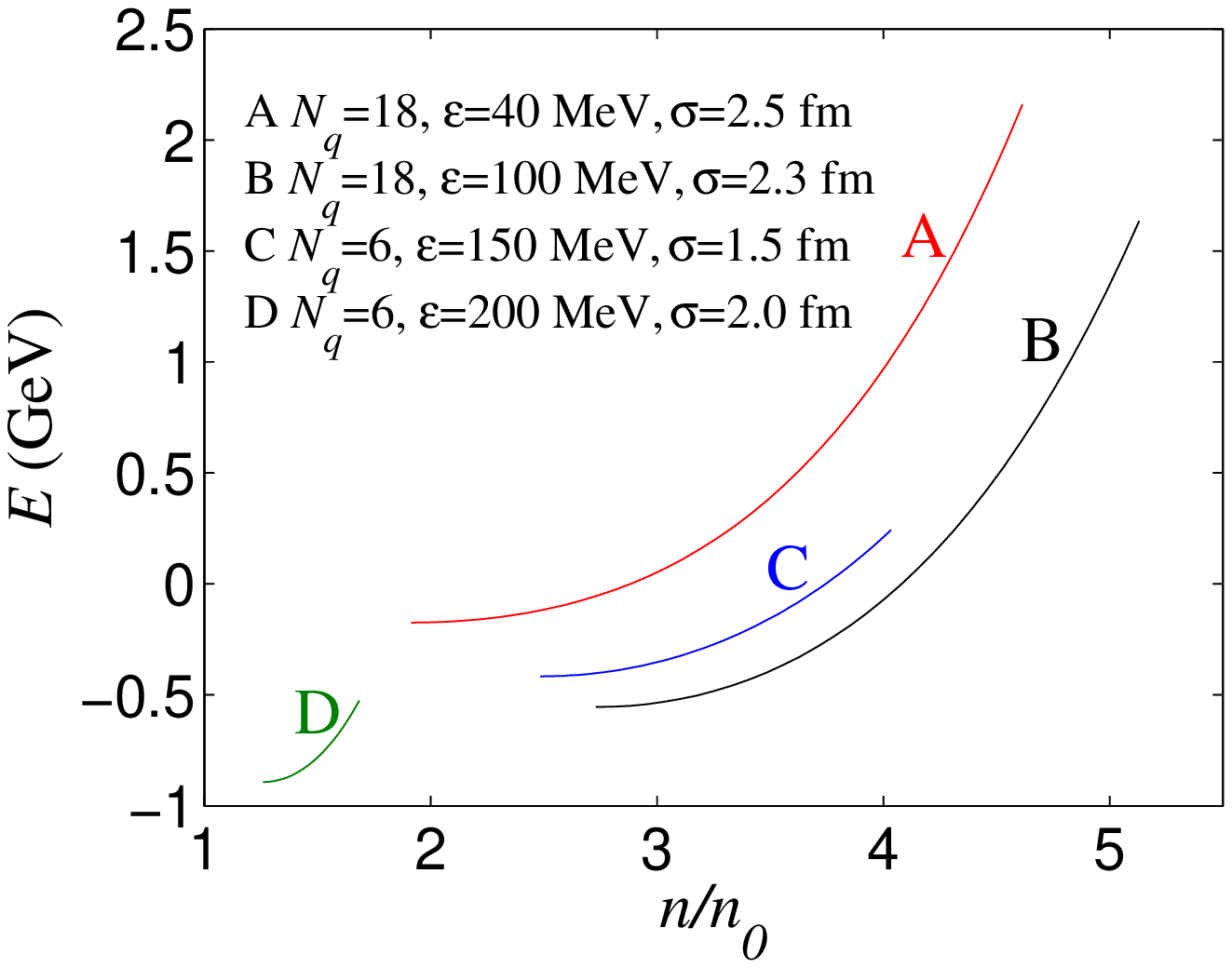}
  \figcaption{The internal energy $E$ per cluster for four groups of parameters
  A(red line), B(black line), C(blue line) and D(cyan line).
  The range of density $n$ is from surface density $n_s$ to the highest central
  density where the quark-cluster stars reaches the
  maximum mass.}\label{f-E}
\end{center}

The mass density $\rho$ consists of rest-mass density and energy density,
\begin{equation}\label{eq-rho}
  \rho=n(N_q m_0+E/c^2),
\end{equation}
then the equation of state for quark-cluster matter can be derived
by combining $P-n$ relation and Eq.~(\ref{eq-rho}), and we show the
results in Fig.~\ref{f-EoS}, for the four groups of parameters A to
D.
\begin{center}
  %\centering
  \includegraphics[width=2.2 in]{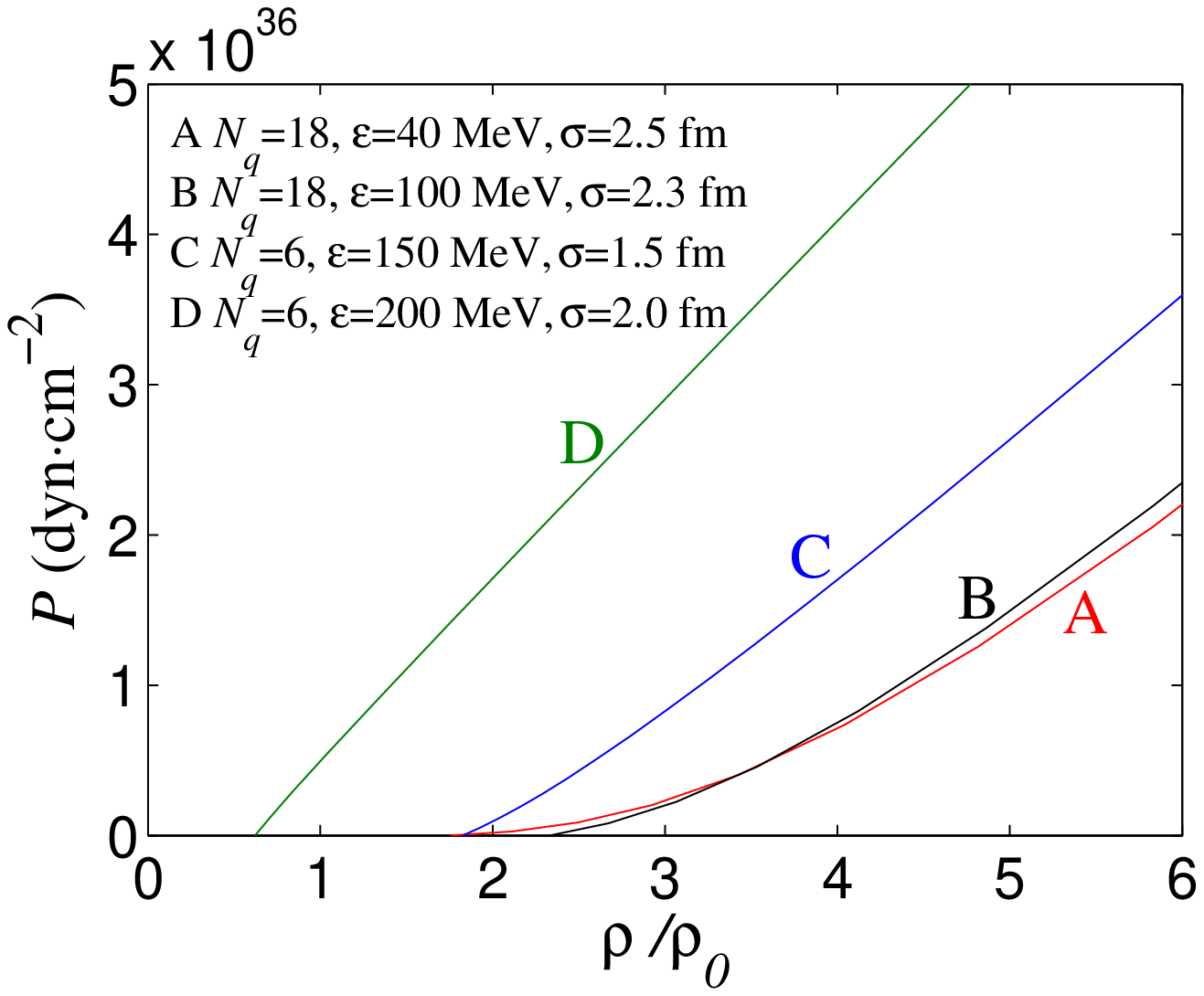}\\
  \figcaption{
  Equations of states for the same four groups of parameters as in Fig.~\ref{f-E}.
  }\label{f-EoS}
\end{center}

\subsection{Mass-radius relation}

Considering perfect fluid case and the general relativity, the hydrostatic
equilibrium in spherically symmetry is described by Tolman-Oppenheimer-Volkoff
equation,
\begin{equation}\label{eq-TOV}
  \frac{dP}{dr}=-\frac{Gm(r)\rho}{r^2}\frac{(1+\frac{P}{\rho c^2})(1+\frac{4\pi r^3P}{m(r)c^2})}{1-\frac{2Gm(r)}{rc^2}},
\end{equation}
where $m(r)=\int_0^r \rho \cdot 4\pi r'^2dr'$.
In the above discussions, we have got the equations of state, from which
we can make a further calculation of the mass-radius and mass-central density
(rest-mass density) relations for quark-cluster stars.
The results are shown in Fig.~\ref{f-m}, for the four groups of parameters A to D,
and we can see that the maximum masses
are higher than three times the solar mass $M_\odot$,
which are reached with central density less than 5$\rho_0$,
for all the selected groups of parameters.
As a comparison, we also plot the mass-radius curves for homogeneous spheres
with the same central density corresponding to each of the four cases.
This shows that the gravity cannot be negligible only when the stars is near the
maximum mass, which could be the result of the strong self-bounding of
quark-cluster stars.
\begin{center}
  %\centering
  \includegraphics[width=2.2 in]{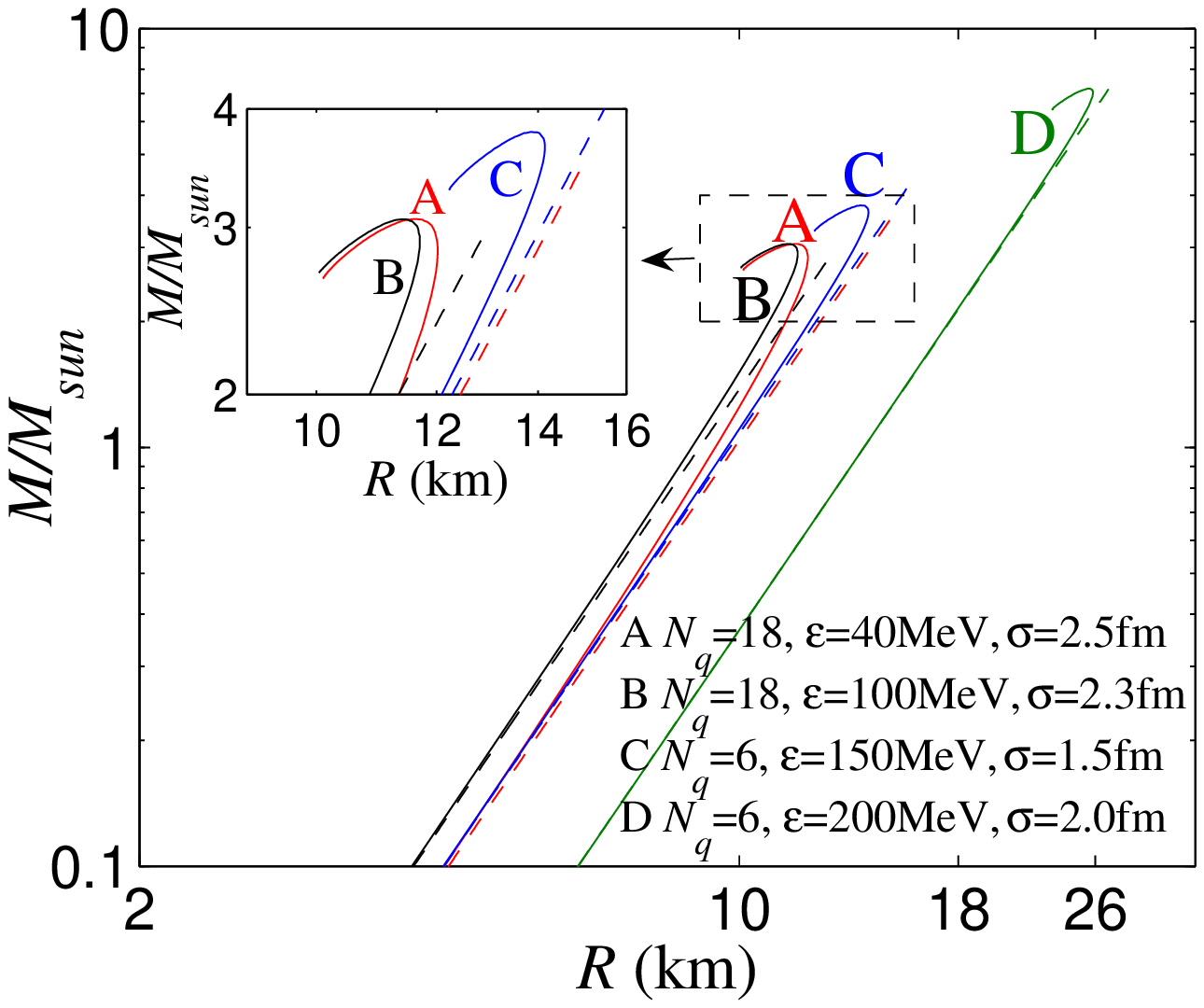}\\
  \includegraphics[width=2.2 in]{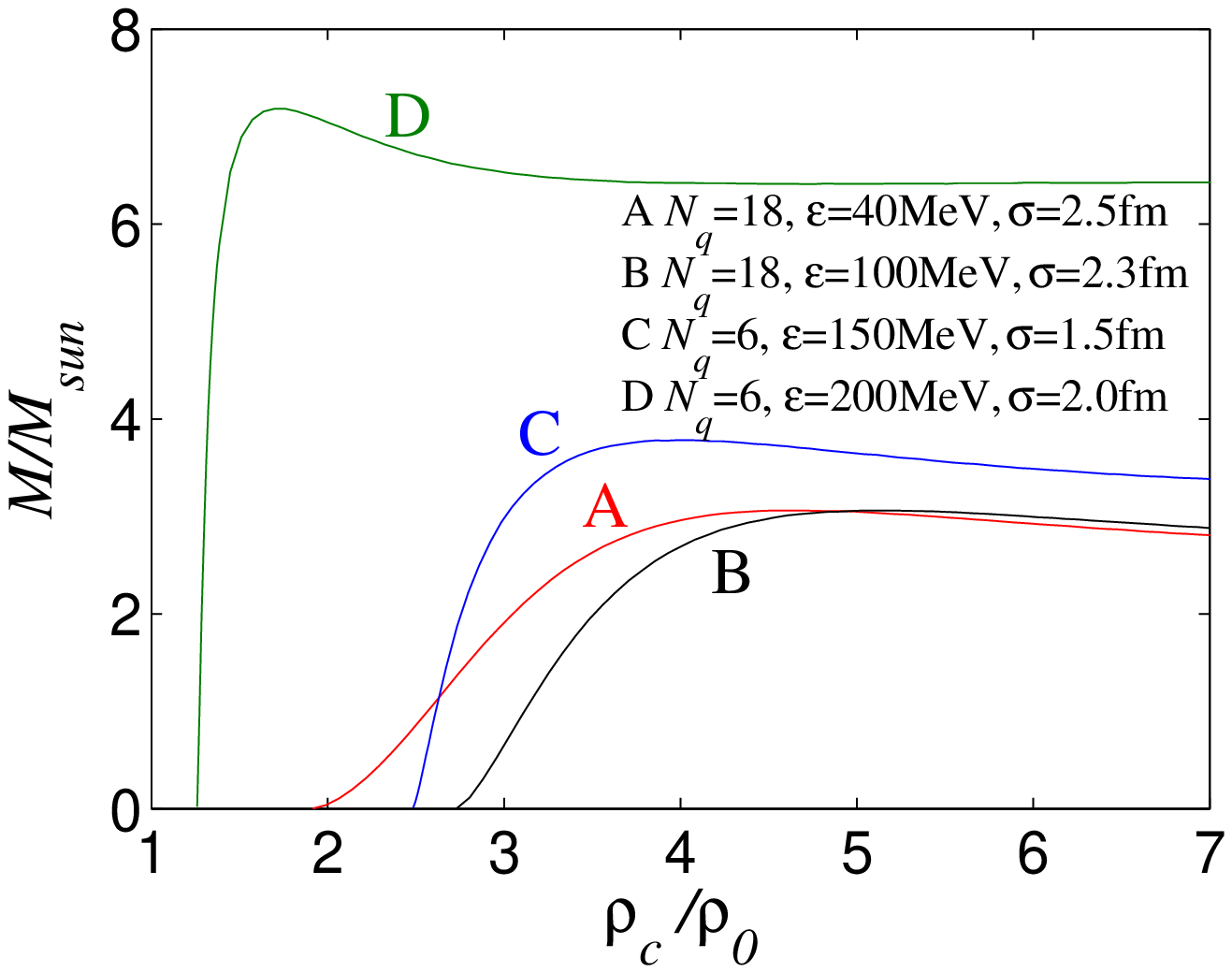}\\
  \figcaption{
  The mass-radius and mass-central density (rest-mass density) curves, and
  different parameters are distinguished by their colors as the same in Fig~\ref{f-E},
  and the corresponding dash lines represent $M=\rho_s \cdot 4\pi R^3/3$.}
  \label{f-m}
\end{center}

Conventional quark matter is characterized by soft equation of state,
and the emerge of quark matter inside compact stars is usually thought
to be a reason for lowering their maximum mass.
The quark-cluster matter, however, could have stiff equation of state
due to the strong coupling.
Although the corresponding-state approach is just a phenomenological and empirical
method, we could still apply it to study the state of quark-cluster matter
and then understand the observations of pulsar-like compact stars.
The observed high-mass pulsar PSR J1614-2230 with mass $1.97\pm0.04M_\odot$
\citep{Demorest:2010bx} has received a lot of attention, and we can see that
the quark-cluster stars in our present model could be consistent with this
observation.
Moreover, our model of quark-cluster stars could not be ruled out even if
the mass of the pulsar J1748-2021B ($2.74M_\odot$) in a galactic cluster is
confirmed in the future.

\section{Melting heat}

If the kinetic energy of quark clusters is much lower than the
inter-cluster potential energy, they may form a solid state which is
meaningful for the thermal X-ray behaviors of compact
stars~\citep{YuM:2011}.
We will estimate the latent heat of phase transition of quark-cluster stars
from liquid to solid state by the corresponding-state approach.

We calculate the ratio of melting heat per particle $H$ and $\varepsilon$ for some
ordinary substances~\citep{CRC}, and find that there is also a good relation between
$H^*=H/\varepsilon$ and $\Lambda^*$, as shown in Fig.~\ref{f-mh1}.
The fitted formula for $H^*$ and $\Lambda^*$ is
\begin{equation}\label{eq-H}
   H^*=1.18e^{-((\Lambda^*-0.12)/1.60)^2}.
\end{equation}
For given $N_q$, $\varepsilon$ and $\sigma$, we can determine $\Lambda^*$
first and then get the value of $H^*$ from Eq.~(\ref{eq-H}), thus the melting
heat $H=H^*\varepsilon$ can be derived.

\begin{center}
  %%\centering
  \includegraphics[width=2.5 in]{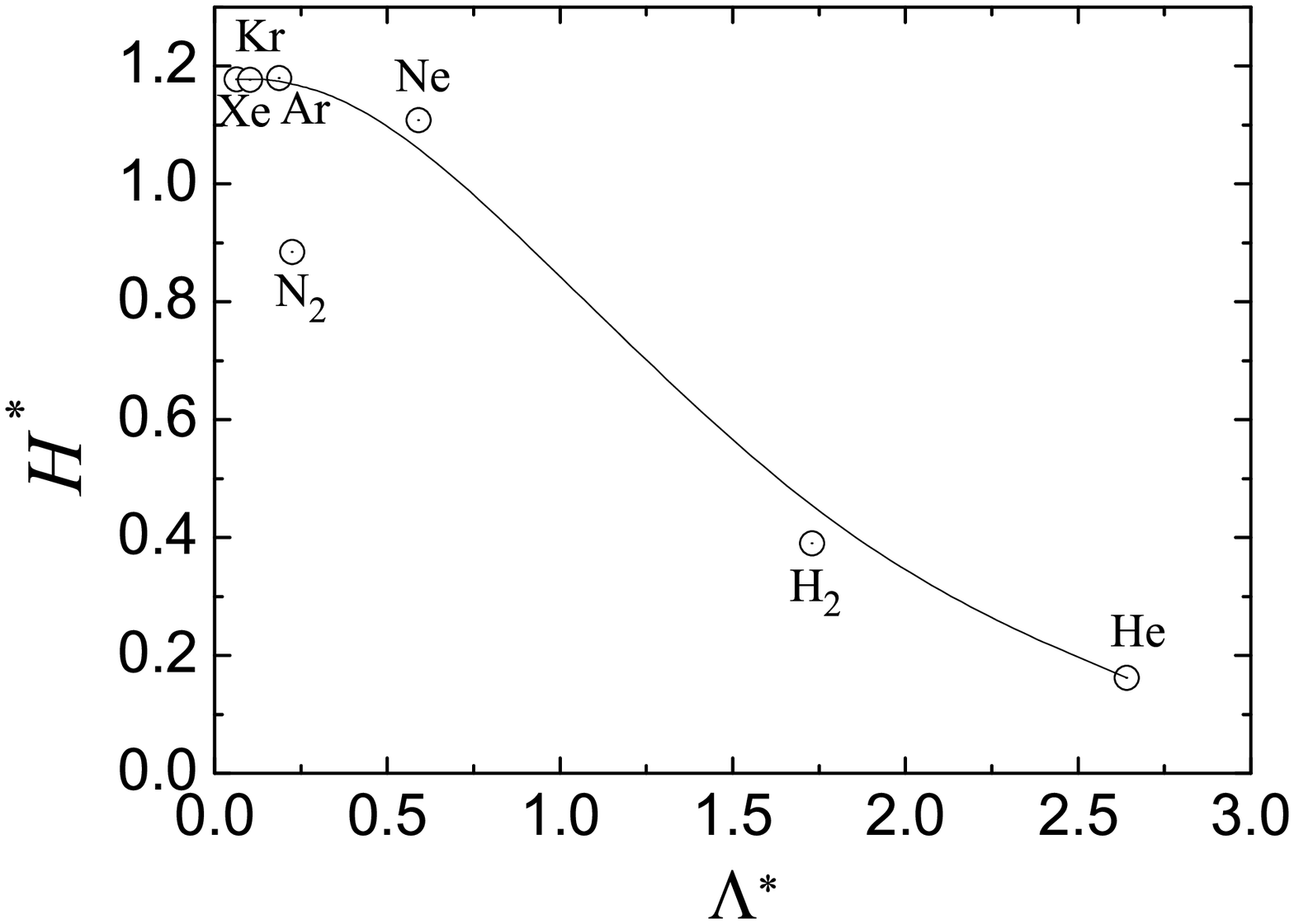}\\
  \figcaption{Data points: experimental data of the reduced melting heat
  $H^*=H/\varepsilon$. Solid line: fitted curve of Eq.~(\ref{eq-H}).}\label{f-mh1}
\end{center}

In Fig.~\ref{f-mh2} pairs of $\varepsilon$ and $\sigma$ which determine the same
melting heat are plotted,  where values of $H$ are chosen to be 1, 10 and 100 MeV,
in two cases $N_q=6$ and $N_q=18$.
The solidification of quark-cluster stars has been suggested to be
relevant to the plateau of $\gamma$-ray burst \citep{DLX-2010}, and
it is found that if
the energy released by each quark-cluster in the liquid to solid phase
transition is larger than 1 MeV, the total released energy could
 produce the plateau.
We can see that under a wide range of parameters in our model,
the latent heat could be sufficient for this way of understanding the plateau of
$\gamma$-ray burst.

\begin{center}
  %\centering
  \includegraphics[width= 2.2 in]{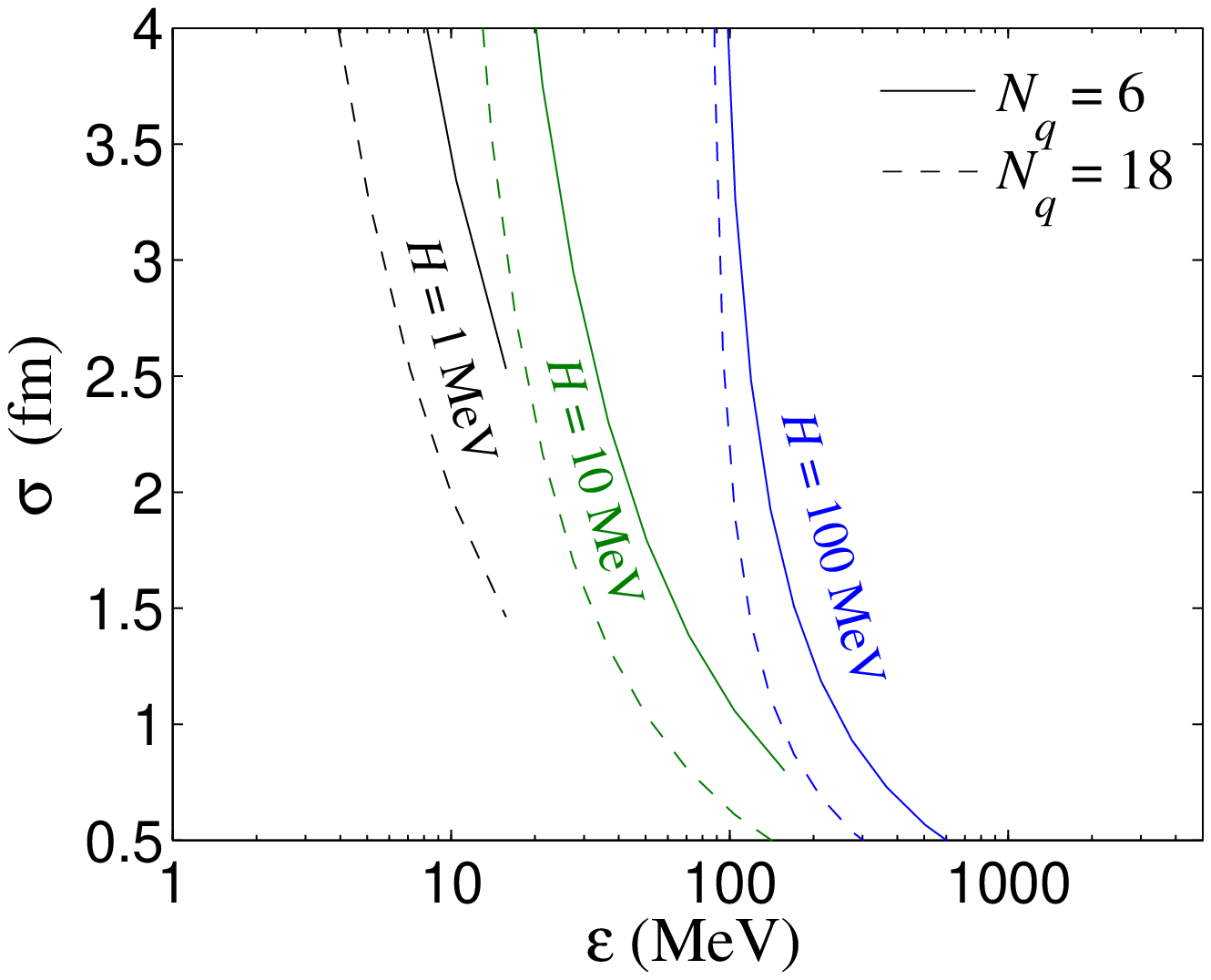}\\
  \figcaption{$\varepsilon$ and $\sigma$ which determine the same
  melting heat of each cluster. $H$ is chosen to be 1 (black line),
  10 (cyan line) and 100 MeV (blue line), in two cases $N_q=6$ (solid line)
  and $N_q=18$ (dashed line).}\label{f-mh2}
\end{center}
%%\centering

\section{Conclusions and discussions}

In cold quark matter at realistic baryon densities of pulsar-like compact stars,
the interaction between quarks would be so strong that they could
condensate in position space, forming quark-clusters, and the stars are then
called quark-cluster stars if the dominant component inside is quark-clusters.
We propose that the interaction between quark-clusters could be
analogous to that between inert gas atoms described by the
Lennard-Jones potential, and apply the corresponding-state approach
to derive the equation of state.
As a phenomenological and empirical method, the corresponding-state
approach can avoid detailed assumptions of quark-cluster matter as
well as computation of the many-body effects, and we only need to
concern about differences between substances.
Along with of these advantages, there are large uncertainty in our results, coming
from the Lennard-Jones approximation and lack of experimental data source.
Even so, the corresponding-state approach could give us qualitative
information about the properties of quark-cluster matter, while the
exact approach under QCD calculations seems to be very difficult and
even impossible now due to significant non-perturbative effects.
Summarily, our two-parameter ($\varepsilon$ and $\sigma$) empirical
approach make it possible to establish a model which could be tested
by observations.

The equation of state we have derived by the corresponding-state
approach could be stiff enough to make a star stable even if its
mass is higher than $2M_\odot$, under reasonable parameters.
This result is consistent with the recent observation of a high-mass pulsar,
thus the emergence of such kind of exotic matter, ``quark-cluster matter'',
could not be ruled out.
The observations of pulsars with higher mass, e.g. $>3M_\odot$, would
even be a support to our quark-cluster star model, and give further constraints
to the parameters.
Moreover, the latent heat released by the solidification of newly born
quark-cluster stars could help us to understand the formation of the
plateau of $\gamma$-ray burst.

Certainly, whether quark-cluster matter could exist at supra-nuclear
densities, and what quark-clusters are composed of, as well as how
to describe their interaction are still open questions.
On the other hand, the nature of pulsar-like compact stars is still
essentially related to significant non-perturbative effects of QCD,
and we hope that future astrophysical observations, complementary to
the terrestrial experiments, could give us hints to all of these
problems.
\\

\acknowledgments{We would like to thank useful discussions at our pulsar group of PKU.}

\end{multicols}

\vspace{10mm}

\begin{multicols}{2}

%\bibitem{Jaikumar06}
%P.~Jaikumar, S.~Reddy, and A.~W.~Steiner 2006
%{\it Phys.\ Rev.\ Lett. } {\bf 96} 041101

\end{multicols}

\clearpage

\end{CJK*}
\end{document}